\documentclass[12pt,preprint]{aastex}

\slugcomment{}

\shorttitle{Fluid Mechanics of Structure Formation}
\shortauthors{Gibson}

\begin{document}


\title{The Fluid Mechanics of Gravitational Structure Formation}


\author{Carl H. Gibson\altaffilmark{1}}
\affil{Departments of Mechanical and Aerospace Engineering and 
Scripps Institution
of Oceanography, University of
California,
     San Diego, CA 92093-0411}

\email{cgibson@ucsd.edu}


\altaffiltext{1}{Center for Astrophysics and Space Sciences, UCSD}


\begin{abstract}

The standard  model for gravitational
structure formation in astrophysics, astronomy, and cosmology  is questioned.  Cold
dark matter (CDM) hierarchical clustering cosmology neglects particle collisions, viscosity,
turbulence and diffusion and makes predictions in  conflict with observations. 
From Jeans 1902 and CDMHC, the non-baryonic dark matter NBDM   forms small clumps
 during the plasma epoch after the big
bang that ``cluster'' into larger clumps.  
 CDM halo clusters collect the baryonic matter (H and He) by gravity so that after 300 Myr of ``dark
 ages'', huge, explosive (Population III) first stars appear, and then galaxies and galaxy
  clusters.   Contrary to CDMHC cosmology,  ``hydro-gravitational-dynamics'' HGD cosmology
suggests the diffusive NBDM material cannot clump and the clumps cannot cluster.  From HGD, the
big bang  results from an exothermic turbulent instability at Planck scales ($10^{-35}$ m).
Turbulent stresses cause an inflation of space and fossil density turbulence remnants that 
trigger gravitational instability at protosupercluster masses
($10^{46}$ kg) in the H-He plasma.  These fragment along plasma
turbulence vortex lines to form protogalaxy masses ($10^{42}$ kg) just
before the transition to gas.  The gas has $\times 10^{-13}$  smaller
viscosity, so it fragments at planetary and globular-star-cluster masses
($10^{25}$ and $10^{36}$ kg) to form the baryonic dark matter (BDM).  
Observations from
the Hubble Space Telescope show protogalaxies (PGs)  in linear clusters 
 reflecting their likely
fragmentation  on plasma  vortex lines.  From merging BDM planets, these PGs gently form
 small stars in globular clusters $\le$1 Myr after the big bang without the
 dark ages, superstars, or reionization  of CDM cosmology.

\end{abstract}


\keywords{turbulence: big bang, fossil---cosmology: theory, observations ---galaxies: dark
matter, star formation}


\section{Introduction}

The standard cold dark matter hierarchical clustering cosmology (CDMHCC) 
conflicts with fluid mechanics theory, and with astronomical observations.
CDMHCC is based on  the Jeans acoustic criterion for gravitational structure formation \citep{jns02,jns29}.  
According to the Jeans theory, density fluctuations $\delta \rho$ on scales smaller than the Jeans
 length $L_J = (V_S/\rho G)^{1/2}$ are stabilized by ``pressure support'', where $G$ is 
 Newton's gravitational constant and $V_S$ is the speed of sound in the gas or plasma.  
 Neither the Jeans theory nor the pressure support justification  withstand scrutiny by  
 more modern fluid 
 mechanical methods that include viscosity, diffusivity, turbulence,
 turbulent mixing, stratification effects and fossil turbulence \citep{gib96}.  Concepts of
``collisionless fluid mechanics'' developed to support CDMHCC \citep{bin87} are also questionable,
from a similar set of questionable assumptions,
and are in conflict with observations that show strong frictional effects exist in galaxy 
interactions \citep{gs03}          
and that the baryonic dark matter of galaxies consists of Jeans-mass 
clumps of earth-mass planets \citep{sch96, gib96}. 

In the hot plasma epoch of the early universe less than 300 000 years ($t =10^{13}$ s) after the big bang, 
the speed of sound $V_S =c/3^{1/3}$ is of order the speed of light $c$.  No plasma structure can
form in the plasma epoch by the Jeans acoustic criterion because the Jeans length scale
exceeds the horizon scale,  $L_J \ge L_H = ct$, where $L_H$ is the  scale of causal connection.  
Information about 
density fluctuations on scales larger than $L_H$ cannot be transmitted in time to produce gravitational
instability because  information speed cannot exceed light speed according to  Einstein's  theory.  
However, observations of temperature anisotropies in the cosmic microwave 
background clearly show well developed
 gravitational structures existed at the $t =10^{13}$ s time of the plasma to gas transition \citep{gib04,gib05}. 

The concept of ``cold dark matter'' was invented to resolve the dilemma.  It was assumed that 
a massive population of virtually collisionless  non-baryonic material was somehow
 created during the nucleosynthesis epoch with 
particle speeds less than $c$.   This ``CDM'' material
condensed into stable clumps, or ``halos'', with a mass determined by the
Jeans scale and  density at its time of creation.  From CDMHC, CDM halos
are gravitationally bound and merge with each other to form larger and larger halos in the process
of ``hierarchical clustering''.  From HGD, CDM halos are diffusively unstable and would be shredded
into particles by tidal forces of a clustering event \citep{gib06}.

From CDMHC, these massive CDM halos then collect the baryonic 
matter into their growing
gravitational potential wells and force the formation of stars, galaxies, galaxy clusters, and eventually
superclusters of galaxies, in that order.  The acoustic peak in the cosmic microwave background spectrum
is often cited as evidence of primordial plasma oscillations in CDM halo wells.  From HGD,
the CMB acoustic peaks  reflect the sonic speed limit of rarefaction waves produced
by gravitational fragmentation of the viscous, expanding plasma as protosupercluster voids formed at density minima. 

The Jeans 1902 acoustic criterion for gravitational instability is flawed by 
a variety of unwarranted assumptions.  Jeans assumed 
self-gravitational condensation of
a collisionless, inviscid, ideal, fluid governed by Euler's momentum equations with Vlasov gravity, neglecting
nonlinear effects and any effects of diffusion.  Even the fluid density was neglected 
to get a solution to these oversimplified equations
  in the ``Jeans' Swindle'' (Binney \& Tremain 1987, p287).  All these assumptions are highly inappropriate for the
expanding superviscous primordial plasma and superdiffusive non-baryonic materials produced by the big bang.  These flaws 
have been pointed out and corrected in a series of papers \citep{gib96,gib00,gib04,gib05} leading to
a fluid mechanical model for gravitational structure formation and cosmology  
termed ``hydro-gravitational-dynamics'' or HGD \citep{gib06}.

According to HGD, gravitational structure formation is guided by viscous, turbulent, and 
diffusional  fluid mechanics, stratified turbulence, and rarely by acoustics.  The big bang
is identified as the first turbulent combustion \citep{gib05}.  Inflation is driven by big bang turbulent
Reynolds stresses, and produces the first fossil turbulence \citep{gib04}, which seeds nucleosynthesis
and provides density fluctuations to seed the first formation of gravitational structures by
fragmentation at density minima.  The first structures are protosuperclusters starting
at $t=10^{12}$ s when the viscous Schwarz scale matches the horizon scale \citep{gib96}.

New observations in a widening range of frequency bands with spectacular 
improvements in spatial and temporal resolution from ground and space telescopes give
a rising flood of contradictions with CDMHCD that can be easily explained or were predicted
 by HGD cosmology.

In the following, HGD cosmology is reviewed  in $\S 2$.   Comparisons to observations are discussed
in $\S 3$.  Finally, conclusions are presented, $\S 3$.

\section{Hydro-Gravitational-Dynamics Theory}

The hydro-gravitational-dynamics theory of gravitational structure formation 
covers a wide range of length scales (Table 1) from the big bang 
Planck scale $L_P = 1.62 \times 10^{-35}$ m of quantum gravitational
instability to the present horizon scale $L_H \approx 10^{26}$ m. The
Planck temperature $T_P = [c^5 h G^{-1} k^{-2}]^{1/2} = 1.40 \times 10^{32}$ K is
so large that turbulence and turbulent mixing are needed to produce  entropy
and make the process irreversible.  

Only Planck particles and Planck anti-particles can exist at such temperatures,
plus their spinning combinations (Planck-Kerr particles), so the viscosity is low
($5 \times 10^{-27}$ m$^2 $s$^{-1}$) and the
Reynolds number is high ($\le 10^{6}$).  
Planck-Kerr particles are the smallest possible Kerr (spinning) black holes. They
represent the big bang
 equivalent of  positronium particles formed from electrons and positrons 
 during the pair production process that occurs at $10^9$ K supernova temperatures.
   Prograde accretion of Planck particles by Planck-Kerr particles can
release up to 42\% of the Planck particle rest mass $m_P = [chG^{-1}]^{1/2}$, 
resulting in the highly efficient exothermic production of turbulent 
Planck gas  \citep{gib04,gib05}.   Large negative turbulent Reynolds stresses 
$\tau_P = [c^{13}h^{-3}G^{-3}]^{1/2} = 2.1\times 10^{121}$ $\rm m^{-1}s^{-2}$ rapidly
stretch space until the turbulent fireball cools from  $T_{P} = 1.4 \times 10^{32}$ K to the strong force freeze-out
temperature $T_{SF} \approx 10^{28}$ K so that quarks and gluons can form.  Besides Planck
particles, the smallest possible Schwarzschild (non-spinning) black hole, only
magnetic monopole particles are possible in the big bang turbulence temperature range  \citep{gut97}.

Gluon viscosity damps the big bang turbulence and increases negative stresses and
the rate of expansion of space.  Turbulence and viscous stresses combine with false vacuum energy in
the stress energy tensor of Einstein's equations to produce
an exponential expansion of space  (inflation) by a factor
 of about $10^{25}$ in the time range $t=10^{-35} \-- 10^{-33}$ s  \citep{gut97}.  Fossil 
temperature turbulence patterns produced by the big bang and  preserved by nucleosynthesis
and cosmic microwave background temperature anisotropies indicate a similar large
value ($10^{5}$) for the big bang turbulence Reynolds number (Bershadski and Sreenivasan, personal
communication 2005).  Only small, transitional Reynolds numbers $ c^{2}t/\nu \approx 10^{2}$ are permitted by photon
kinematic viscosity $\nu = 4 \times 10^{26}$ $\rm m^2 s^{-1}$ at the time ($t=10^{12}$ s) of first structure \citep{gib00}.

Gluon, neutrino, and photon viscosities dominated momentum transport and 
prevented turbulence during the electroweak, nucleosynthesis, and energy 
dominated epochs before $t=10^{11}$ s, and also the formation of gravitational 
structures.  Soon after this beginning of the matter dominated
epoch the neutrinos ceased scattering on electrons and became super diffusive.  Neutrinos
were produced in great quantities at the $10^{-12}$ s electroweak transition that may still exist
as part, or most, of the non-baryonic dark matter that dominates the mass of the universe.
Momentum transport became dominated by photon viscosity, with the possibility of weak turbulence
\citep{gib00,gib06}.

The conservation of momentum equations for a fluid subject to viscous, magnetic and other forces is
\begin{equation}
{\partial  \vec{v} \over \partial t }= -\nabla  B + \vec{v} \times \vec{\omega}
+ \vec{F_g} + \vec{F_\nu} + \vec{F_m} + \vec{F_{etc.}}
\end{equation}
where $B=p/\rho + {v^2}/2$ is the Bernoulli group, $p$ is pressure, $\rho$ is density, $\vec v$
is velocity, $t$ is time, $\vec{\omega}$ is vorticity, $\vec{F_m}$ is 
magnetic force, and $\vec{F_{etc.}}$ are miscellaneous other forces.  In the early 
universe, $\nabla  B$, $\vec{F_m}$. 
and $\vec{F_{etc.}}$ are small compared to the 
inertial-vortex force $\vec{v} \times \vec{\omega}$ that causes turbulence
and the viscous force $\vec{F_\nu}$ and gravitational  force $\vec{F_g}$.

When viscous and turbulence forces as well as diffusion are taken into account, the HGD cosmological
criterion for gravitational instability at scale $L$ becomes
\begin{equation}
L_H \ge L \ge L_{{SX}_{max}} = max[{L_{SV},L_{ST},L_{SD}]}
\end{equation}
where $L_H$ and the Schwarz  scales 
$L_{SV},L_{ST},L_{SD}$ are included in summary Table 1.

The initial stages of gravitational instability are very gentle, driven by
either positive or negative density variations $\delta \rho$.  All forces
other than $\vec{F_g}$ on the right hand side of the momentum equations vanish.  
Pressure support cannot prevent gravitational instability because any
forces from enthalpy $p/\rho$ gradients are perfectly balanced by gradients of kinetic
energy ${v^2}/2$  as the fluid starts to move ($\nabla  B= 0$).  Because the universe is
uniformly expanding with rate-of-strain $\gamma \approx t^{-1}$, gravitational 
condensations on density maxima are inhibited by the expansion
but gravitational fragmentations at density minima are enhanced.  The first 
gravitational structures occurred by fragmentation at density minima when the horizon scale $L_H$
increased to exceed the plasma Schwarz scales, which were all 
nearly equal in the low Reynolds number
 hot plasma epoch, $L_{ST} \approx  L_{SV}\approx  L_{SD}$.
 
 In the turbulent primordial plasma, maximum stretching rates and the first
 fragmentation should occur along vortex lines of the turbulence, as shown in
 Figure 1.  The large initial density $\rho _0$ of the plasma is preserved as
 a fossil by the density of the protogalaxies $\rho _{PG} \approx \rho _0$ and
 the protoglobularclusters $\rho _{PGC} \approx \rho _{GC} \approx \rho _0$.

\section{Comparison to Observations}

Observations relevant to the formation of gravitational structures in the early
universe are appearing at a rapid pace from many space satellites and highly
 refined ground based  telescopes.  Figure 2
contrasts the HGD and CDMHC interpretations of the cosmic microwave background
observations of the Wilkinson Microwave Anisotropy Probe (WMAP), from
the NASA WMAP Science Team website (white labels).  According to HGD cosmology
 (magenta and yellow labels), gravitational structures appeared much earlier than
predicted by CDMHC cosmology.  From HGD the first stars formed very gently in dense
globular star clusters immediately after the transition to gas; that is, in less than a
million years.  From CDMHC and the Jeans 1902 theory the first stars were enormous (Mass  $ \ge 110 M_{\sun}$)
and explosive, and could not form for 300 million years to permit sufficient 
cooling from the Jeans criterion.  

The intense starlight from these Population III superstar explosions is excluded
 because it is not observed in 
$\gamma$-ray spectra from blasars \citep{ah06}.  Pop. III starbursts should 
prevent the formation of globular star clusters with their large densities
of small ancient stars, and should have contaminated the entire universe with large atomic number
chemicals.  However, dense molecular clouds form stars with low metalicity, and globular
star clusters have low metalicity as expected from HGD.  

The Pop. III event is supported by a lack of hydrogen gas in quasar spectra
with redshift $z \le 6$ ($t \ge $ 300 Myr) termed the ``Gunn-Peterson trough'' (predicted
in 1965) attributed to reionization of the gas by Pop. III stars, but this observation
can also be explained by HGD cosmology that predicts by this time most of the 
primordial H-He gas should be condensed and frozen as planets in protoglobularstarclusters
and thus undetectable in such quasar spectra.
Evidence of metallically in distant quasars is taken as evidence of Pop. III stars, but is equally
explained by HGD formation of a relatively small number of Pop. III explosive stars 
near protogalaxy centers as the density increases from enhanced rates of PGC and PFP mergers.

Figure 3 shows a Hubble Space Telescope 
Advanced Camera for Surveys (HST/ACS) image of the Tadpole galaxy merger \citep{gs03}.  
One CDMHC cosmology interpretation is that the filamentary galaxy VV29b is a collisionless ``tidal tail'' 
triggered by the close passage of the small galaxy VV29c near the larger galaxy VV29a, which
flings VV29c far into the background  without frictional losses \citep{bri01}. Dust trails show the
VV29cdef fragments have frictionally merged into embedded positions.  This
is confirmed by observations in a variety of infrared bands \citep{jar06}.    Another CDMHC interpretation
\citep{tre01} is that the filamentary galaxy VV29b constituting the tail of the Tadpole has revealed
an invisible CDM halo object.  Frictionless tidal tail models \citep{tt72} along 
with frictionless stellar and galactic dynamics models \citep{bin87} conflict with observations and must be abandoned.

The HGD interpretation is that
the spiral paths around VV29a are frictional accretion paths of galaxy fragments VV29cdef revealed
by star formation in the VV29a baryonic dark matter halo.  This interpretation is strongly supported
by the 42-48 young globular star clusters arranged precisely in a row pointing to the point of
 frictional capture of VV29c, and by the high resolution HST images that show VV29c is embedded
 in VV29a, not far in the back.  The extent of the BDM halo of VV29a is shown to be
 $4 \times 10^{21}$ m (130 kpc) by the Tadpole merger system.  This indicates a frozen
 PGC diffusivity of about $10^{27}$ m$^2$s$^{-1}$.
 
 The inset in Fig. 3 lower left shows an example of the chain galaxies visible in the Tadpole image
 (circle) of a chain clump cluster \citep{elm05a, elm05b}.  From HGD cosmology, this
  linear configuration of the dimmest galaxies of the HST ultra-deep-field supports the idea
   that galaxies were formed by fragmentation along turbulent vortex lines in the
    plasma epoch, as shown in Fig. 1.  As the PGCs freeze they diffuse into BDM halos with
    size $L_{SD} \approx [Dt]^{1/2}$.  The luminous halos of the chains of clumps are
    interpreted as stars and YGCs triggered into existence
     by tidal forces of the protogalaxy separations.

\section{Conclusions}

The standard CDMHC cosmology, the Jeans 1902 criterion for gravitational structure formation
and the ``collisionless'' concepts of galactic dynamics \citep{bin87}  and frictionless tidal tail
formation \citep{tt72} are in
fundamental conflict with modern fluid mechanics and make predictions that are increasingly
in conflict with observations, eg. \citep{ah06} and Fig. 3.  They must be abandoned.  
The Jeans theory and CDMHC cosmology
 have been modified and extended  to include important effects of quantum gravity, viscosity, turbulence,
fossil turbulence and diffusion on gravitational structure formation and cosmology in a
new paradigm termed hydro-gravitational-dynamics, or HGD \citep{gib96, gib00, gib04, gib05, gib06}.  

The predictions of HGD about the formation of structure are very different from those
of CDMHC, as shown in Fig. 1 and Fig. 2.  From HGD, the largest structures form first rather 
than last as predicted by CDMHC.  Galaxies and clusters of galaxies emerge from the plasma
epoch in linear morphologies reflecting the weak plasma turbulence that triggered their 
fragmentation, as shown by observations in Fig. 3.  The gas universe turned to a fog
of planetary mass clouds (PFPs) in protoglobularstarcluster clumps (PGCs) that persist as the baryonic
dark matter (BDM), as observed from quasar microlensing \citep{sch96}.
Because all stars form from planets in dense clumps according to HGD, it is easy to explain
claims of dark energy and accelerating rates of the universe expansion as due to dimming 
effects of the BDM.  Linear strings of stars and young globular clusters of stars are shown
in the Tadpole system in Fig. 3 that clearly demonstrate the frictional accretion of galaxy fragments,
and the large extent of the BDM halo formed by frozen PGCs diffusing away 
from the original dense protogalaxy.  

Chain galaxies detected in the HST/ACS Tadpole
images, Fig. 3, are interpreted as chains of protogalaxies formed by gravitational
fragmentation along turbulent vortex lines
in the last stages of the plasma epoch according to HGD, Fig. 1.  Luminous
halos of the massive star clumps reflect star formation by tidal
forces of the separation triggered in the dark PGCs of the BDM protogalaxy halos.

\acknowledgments


\begin{figure}
        \epsscale{1.2}
        \plotone{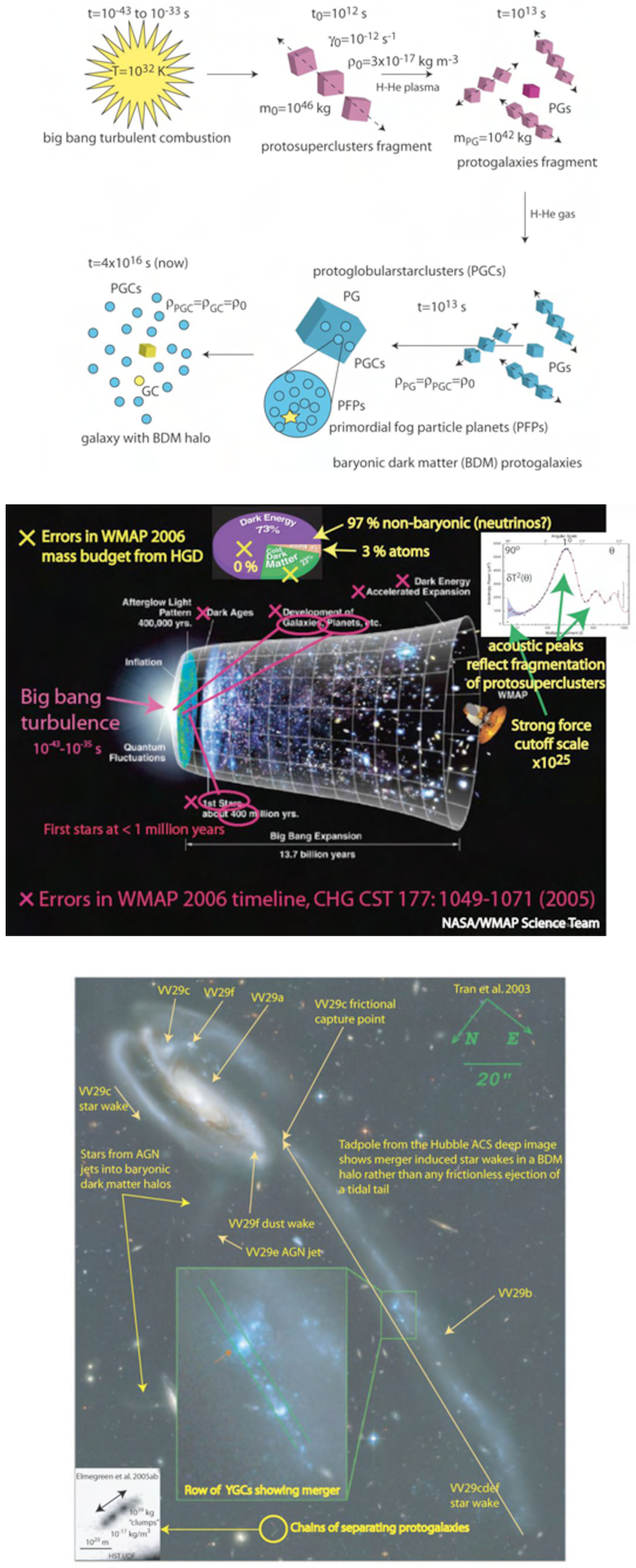}
        \caption{ }
      
       \end{figure}


      (top) 1.  Formation of gravitational structure according to hydro-gravitational-dynamics (HGD)
         theory (modification of Fig. 1 in Gibson 2006).  
        A turbulent instability causes the big bang \citep{gib04,gib05}, upper left.  Fossil temperature
        turbulence fluctuations produce density patterns that trigger gravitational fragmentation along
        turbulent vortex lines at protosupercluster scales, top center, and protogalaxy scales, 
        top right \citep{gib96,gib00}.   When the plasma turns to gas the viscosity and fragmentation 
        scales decrease, top and bottom right,  causing fragmentation at PGC globular cluster and PFP 
        planet scales to form the BDM and the first small stars, bottom center.  The dense, dark, PGCs 
        have now diffused to form large dark matter galaxy halos, bottom left. 
        
       (middle) 2. NASA/WMAP Science Team  mass budget (top) and timeline  
        for gravitational structure formation 
        (white captions) based on
        the standard CDMHC cosmology,   compared to predictions
        (yellow and magenta captions) from  HGD cosmology  \citep{gib96, gib00, gib04, gib05}.
        The acoustic peaks in the anisotropy power spectrum (upper right) are interpreted
        as a reflection of protosupercluster fragmentation, as shown in Fig. 1. 
        
        (bottom) 3. The ``Tadpole'' galaxy reveals the frictional  baryonic dark matter halo of
        galaxy VV29a by the star and young globular cluster wakes produced as
        galaxy fragments VV29cdef merge forming the filamentary galaxy VV29b \citep{gs03}.  Chains
        of proto-galaxies (insert lower left) reflect their gravitational fragmentation along 
         turbulent vortex lines of the plasma.  The glow reveals frictional
         formation of YGCs in the baryonic dark matter halos of the PGs as they separate in the gas
         epoch due to the expansion of the universe \citep{gib06}. 
       
       \begin{deluxetable}{lrrrrcrrrrr}
\tablewidth{0pt}
\tablecaption{Length scales of self-gravitational structure formation}
\tablehead{
\colhead{Length scale name}& \colhead{Symbol}           &
\colhead{Definition$^a$}      &
\colhead{Physical significance$^b$}           }
\startdata 
Compton wavelength & $L_{C}$&$ h m ^{-1} c ^{-1}$& wavelength of particle with mass $m$

\\Schwarzschild radius & $L_{S}$ & $ G m c^{-2}$ & size of non-spinning black hole of mass $m$

\\Planck length & $L_P$ & $[c^{-3}hG]^{1/2}$ & scale of big bang vacuum instability

\\Jeans Acoustic & $L_J$ &$V_S /[\rho G]^{1/2}$& ideal gas pressure
equilibration
\\ Schwarz Diffusive & $L_{SD}$&$[D^2 /\rho G]^{1/4}$& $V_D$ balances $V_{G}$
\\  Schwarz Viscous & $L_{SV}$&$[\gamma \nu /\rho G]^{1/2}$& viscous force
balances gravitational force
  \\ Schwarz Turbulent & $L_{ST}$&$\varepsilon ^{1/2}/ [\rho G]^{3/4}$&
turbulence force  balances gravitational force
\\

Kolmogorov Viscous & $L_{K}$&$ [\nu ^3/ \varepsilon]^{1/4}$& turbulence
force  balances viscous force
\\

Ozmidov Buoyancy & $L_{R}$&$[\varepsilon/N^3]^{1/2}$& buoyancy force
balances turbulence force
\\

Particle Collision & $L_{Col.}$&$ m \sigma ^{-1} \rho ^{-1}$& distance between
particle collisions
\\

Hubble Horizon & $L_{H}$&$ ct$& maximum scale of causal connection
\\


\enddata
\tablenotetext{a}{$V_S$ is sound speed, $\rho$ is density, 
$D$ is the diffusivity, $V_D \equiv D/L$ is the diffusive velocity
at scale $L$, $V_G \equiv L[\rho G]^{1/2}$ is the gravitational velocity,
$\gamma$ is the strain rate,
$\nu$ is the kinematic viscosity,
$\varepsilon$ is the viscous dissipation rate, $N \equiv
[g\rho^{-1}\partial\rho/\partial z]^{1/2}$ is the stratification frequency,
$g$ is self-gravitational acceleration, $z$ is in the opposite direction
(up),
$m$ is the particle mass,
$\sigma$ is the collision cross section,  $c$ is light speed, $h$ is 
Planck's constant,  $G$ is Newton's
constant,  $t$ is the age of
universe.}

\tablenotetext{b}{Magnetic and other forces (besides viscous and turbulence)
are negligible for the epoch of primordial self-gravitational structure
formation
\citep{gib96}.}


\end{deluxetable}

\end{document}